\documentstyle[psfig,prl,aps,epsf]{revtex}

\begin{document}
\draft

\title{Classical Correspondence of Unruh Effect}
\author{Shih-Yuin Lin\footnote{
Electronic address: {\tt sylin@phys.sinica.edu.tw}}}
\address{Institute of Physics, Academia Sinica, Nankang, Taipei 11529,
TAIWAN}
\date{April 2001}
\maketitle

\begin{abstract}
Derived from the semi-classical quantum field theory in curved spacetime,
Unruh effect was known as a quantum effect. We find that there does exist a
classical correspondence of this effect in electrodynamics. The thermal
nature of the vacuum in correlation function for the uniformly accelerated
detector is coming from the non-linear relationship between the proper time
and the propagating length of the electromagnetic wave. Both the Coulomb
field of the detector itself and the radiation supporting the detector's
uniformly accelerating motion contribute to the non-vanishing vacuum
energy. From this observation we conclude that Unruh temperature
experienced by a uniformly accelerated classical electron has no additional
effects to Born's solution for laboratory observers far away from the
classical electron.

\end{abstract}

%\pacs{PACS numbers: }

A uniformly accelerated particle moving in Minkowski vacuum is claimed to
see a thermal bath with a temperature proportional to its proper
acceleration $a$. This was first derived by Davies using the quantum field
theory in curved spacetime, then a realization by a model with a uniformly
accelerated particle detector was established by Unruh\cite{unruh}. In
Unruh's model, the particle detector is an idealized point-like object
moving in a classical worldline $x'(\tau)$, with the detector-field
interaction described by the interacting action
\begin{equation}
  S_{\rm int} = e\int d\tau d^4 x\sqrt{-g} q(\tau )\phi (x)
    \delta^4 (x-x'(\tau)),\label{Sint}
\end{equation}
where $q(\tau)$ is the monopole moment of the detector, $e$ is the
coupling constant, and $\phi$ is the scalar field to be detected.
Suppose the detector was prepared in its ground state at past null
infinity $(\tau\to -\infty)$. Then, for sufficiently small $e$, the
transition rate may be given by the first perturbation theory as\cite{BD}
\begin{eqnarray}
P&=&{e^2\over 2\pi}\sum_{E\not=E_0}\left|\left< E|q(0)|E_0\right>
  \right|^2\int^\infty_{-\infty}d(\Delta\tau) e^{-i(E-E_0)(\Delta\tau)}
  D^{(1)}(\Delta\tau)\nonumber\\
&=& {e^2\over 2\pi}\sum_{E\not= E_0} {(E-E_0)\left|\left< E|q(0)
  |E_0\right>\right|^2\over e^{2\pi(E-E_0)/a}-1}+{\rm singular\ terms}.
\label{probab}
\end{eqnarray}
The singular terms in above transition rate are owing to the point-like
property of the detector, and would be subtracted by some reasonable
renormalization schemes. Eventually the thermal character could be
recognized by extracting the Planck factor in the finite part of
Eq.$(\ref{probab})$. One can further prepare the detector in equilibrium 
with the thermal bath initially, then the expectation value of the detector 
energy would be static with a Planckian-like spectrum.

Unruh effect was thought of as a pure quantum phenomenon. Nevertheless, 
since the quantum correlation function responsible for the Planck factor in 
Eq.$(\ref{probab})$ is also a Green's function in corresponding classical 
field theory, if there exists thermal characters in a semi-classical 
theory, similar informations should be found in its classical counterpart.
Boyer\cite{boyer}, for example, had illustrated a classical version of 
Unruh effect: if one introduces random phases in the mode expansion of 
classical fields then averages them out, one can obtain similar effects. 
However, the random phases can be considered as a substitute of the 
zero-point fluctuation, thus an outsider for the classical field theory. 
With Boyer's result one still cannot give a definite answer that Unruh 
effect is essentially quantum or classical.

So far the attempts at observing the Unruh effect in laboratories are
mainly focused on measuring the responses of accelerated
electrons\cite{bell}. The motivations of these proposals are reasonable
when one notes that the classical relativistic electron theory is similar
to Unruh's detector theory in the structure of the interacting
action\cite{lin0106}. Actually, in classical electrodynamics, the uniformly
accelerated charge(UAC) had been an interesting problem for a long time,
though it is impossible to prepare any perfect experiment of this kind in a
laboratory. One may take the advantage of an extensive literature about
classical UAC in studying (possibly) related topics to Unruh effect.

The solution of the electromagnetic(EM) field associated with
UAC was first given by Born in 1909\cite{born}. In his solution the
magnetic field vanishes at $t=0$ hypersurface, hence it was claimed that
there is no wave-zone in this system\cite{pauli}. Half a century after,
Bondi and Gold\cite{BG} gave a more general solution satisfying Maxwell
equations in the whole spacetime. Then Fulton and Rohrlich\cite{FR} found
that, actually, the radiating power flux does not vanish on the future
lightcone; rather, it is $2 e^2 a^2/3$ where $e$ is the electric charge of
the testing particle. Finally, Boulware\cite{boul} understood that the UAC
not only radiates but also absorbs EM power. To keep the charge in its
constant acceleration, there has to have a power-input assigned in the
boundary condition at the past null infinity\cite{nester}.

The motion of a charged particle in EM field is described by the
Lorentz-Dirac equation\cite{dirac},
\begin{equation}
  m a^\mu = {e\over c}F_{\rm in}^{\mu\nu}v^\nu +F^{\mu}_{\rm ext}+
  \Gamma^\mu,
\end{equation}
where $m$, $e$, $\tau$ denote the mass, charge and the proper time for the
particle, respectively, $v^\mu\equiv dx^\mu /d\tau$, $a^\mu\equiv
dv^\mu /d\tau$, $F^{\mu}_{\rm ext}$ is the non-EM force, and
\begin{equation}
  \Gamma^\mu \equiv (F^{\mu\nu}_{\rm ret}-F^{\mu\nu}_{\rm adv})v_\nu =
  {2e^2\over 3c^3}\left(\dot{a}^\mu -{1\over c^2}a^\nu a_\nu v^\mu\right)
\end{equation}
is the difference of the radiation from the absorption of the particle.
It is clear that, while the radiating power is measured at future null
infinity globally, the field-strength differences influencing the particle
motion are measured locally.

For a classical charge in uniform acceleration, the difference between the
retarded and advanced field-strengths vanishes, {\it i.e.}, $\Gamma^\mu
=0$. This can be interpreted as the particle emits and absorbs photons in
the same rate. Similarly, the Unruh effect from semi-classical field theory
states that a uniformly accelerated detector in equilibrium with a
thermal bath not only absorbs and counts the photons, but also emits
photons in the same rate.

The trajectory of a uniformly accelerated charge in Minkowski
space with proper acceleration $a$ in z-direction is a hyperbola
in $t$-$z$ plane, namely,
\begin{equation}
  x^\mu = \left((ac)^{-1}\sinh a\tau , a^{-1}\cosh a\tau ,0,0\right),
\end{equation}
where $\tau$ is the proper time in the accelerated charge's
coordinate.\footnote{In this letter, we use the cylindrical coordinate
$d\tau^2 = c^2dt^2-dz^2 -d\rho^2-\rho^2 d\phi^2$.}
The total field strength of the EM field in a half of the
spacetime is given by\cite{born}\cite{BG}
\begin{eqnarray}
  E_z &=& F^{tz}=-{4e\over a^2\xi^3}\left( a^{-2}+c^2t^2-z^2+\rho^2\right)
    \theta(z+ct),\label{bornEz}\\
  E_\rho &=& F^{t\rho} ={8e\rho z\over a^2\xi^3}\theta(z+ct)
    +{2 e\rho\over \rho^2+ a^{-2}}\delta(z+ct),\label{bornEr}\\
  B_\phi &=& F^{z\rho} ={8e\rho t\over a^2\xi^3}\theta(z+ct)
    -{2 e\rho\over \rho^2+ a^{-2}}\delta(z+ct),\label{bornBp}
\end{eqnarray}
where
\begin{equation}
  \xi \equiv\sqrt{4a^{-2}\rho^2+\left( a^{-2}+c^2t^2-z^2-\rho^2\right)^2},
\end{equation}
and the function $\theta (x)$ is defined by
\begin{equation}
  \theta(x)=\left\{\begin{array}{l}1\mbox{ for }x>0\\{1/2}\mbox{ for }
  x=0\\0 \mbox{ for }x<0\end{array}\right. .
\end{equation}
Note that $F^{\mu\nu}$ is not analytic at $z+t=0$.
Suppose an observer comoving with the accelerated charge has the
trajectory\footnote{The location of the observer is chosen such that
the inverse Fourier transformation below exists or physically, the clocks
of the detector and the observer can be synchronized without destroying the 
Lorentz invariance of the whole system.}
\begin{equation}
  x^{\mu}=\left( (ac)^{-1}\sinh a\tau, a^{-1}\cosh a\tau ,\rho,0\right),
  \label{obser}
\end{equation}
with the same $t$ and $z$ as the charge.
Then the classical energy density measured by it is
\begin{eqnarray}
  {\cal E} &\equiv& 4\pi T_{\mu\nu}(\tau)v^\mu (\tau) v^\nu (\tau)
    = {e^2\over 2\rho^4}\left( 1+{a^2\rho^2\over 4}\right)^{-2}
    \nonumber\\&=& {e^2\over 2}\left(
    {1\over\rho^4}-{a^2\over 2\rho^2} + {3a^4\over 16}+O(\rho^2)\right),
\end{eqnarray}
where the stress-energy tensor for EM field is
\begin{equation}
  T^{\mu\nu} = -{1\over 4\pi}\left( F^{\mu\alpha}F^\nu{}_\alpha -
  {1\over 4}g^{\mu\nu}F^{\alpha\beta}F_{\alpha\beta}\right),
\end{equation}
and $v^\mu(\tau)$ is the four-velocity of the charge at its proper time
$\tau$. When the observer gets closer to the accelerated charge, {\it i.e.}
$\rho\to 0$, there are two singular terms present in above small-$\rho$
expansion. \footnote{Here the accelerated charge has $T_{\mu\nu}u^\mu v^\nu
=0$, where $u_\mu$ is any spacelike vector orthogonal to $v^{\mu}$. This
means that the net power-flow is zero, rather than there is no radiation
from this uniformly accelerated charge. The radiated power is simply
balanced by the absorbed power. Hence the radiation energy is non-vanishing
for the accelerated charge.}
%The $\rho^{-4}$-term corresponds to the static Coulomb energy of the
%charge, and the $\rho^{-2}$-term corresponds to the radiation energy.
Interesting enough, the third term is non-vanishing as $\rho\to 0$.

As a conservative quantity, $\cal E$ is a constant of the proper time
$\tau$. Its dependence on time-like variables could be introduced as
follows. By virtue of the Lorentz invariance of the system, it suffices to
study the EM field at the time slice $t=0$ without loss of generality.
When $t=0$ ($\tau=0$), at $\rho$, the total
field strength is the retarded field strength starts at the proper time
$\tau_- =-\Delta/2$ and the advanced field strength ends at $\tau_+ =+
\Delta/2$. This gives a correlation
\begin{equation}
  \rho = \sqrt{\left({1\over a}\sinh {a\over 2}\Delta\right)^2 -
    \left({1\over a}\cosh{a\over 2}\Delta -{1\over a}\right)^2}
    ={2\over a}\sinh {a\over 4}\Delta\label{rhodel}
\end{equation}
between $\rho$ and the parameter $\Delta$ (see FIG.1).
It should be emphasized here that $F_{\mu\nu}(\Delta )$ are the field
values at $t,z=0$ and the $\rho$ in Eq.$(\ref{rhodel})$, rather than the
field values at the position of the point charge with $\tau =\pm\Delta/2$.
Also, $\Delta$ is not a measurable for the apparatus in laboratories.
\begin{figure}
\centerline{\psfig{figure=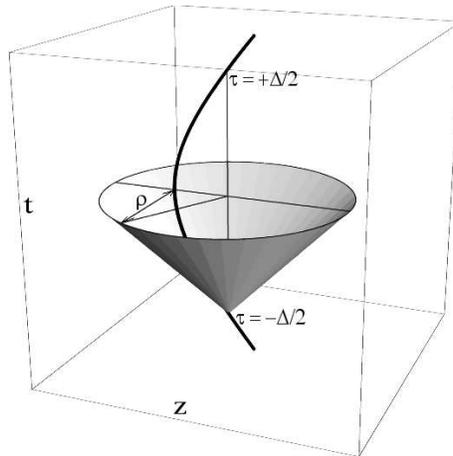,height=6cm,angle=0}}
\caption{The bold line is the hyperbolic trajectory of the UAC in
spacetime. The lightcone starts from the UAC at its proper time $-
\Delta/2$, and reaches the two-sphere of the lightfront at $t=0$
hyper-surface. The latter with radius $a^{-1}$sinh$(a\Delta/2)$ is
represented by a circle in this figure.}
\end{figure}

When $\rho\to 0$, $\Delta\to 0$ also, and we recover the energy density
represented in the well-known autocorrelation function\cite{boyer}(up to an
overall factor proportional to the fine-structure constant $\alpha\sim
e^2/\hbar c$\cite{mane}) as
\begin{eqnarray}
  {\cal E}&=& \lim_{\Delta\to 0}{e^2 a^4\over 2\sinh^4 (a\Delta/2)}
    = \lim_{\Delta\to 0}{e^2\over 2}\left<
    E_i^{\rm ret}(-\Delta/2)E^i_{\rm adv}(\Delta/2)\right>\nonumber\\
    &=& \lim_{\Delta\to 0}{e^2\over 2}\left( {16\over\Delta^4}
-{8 a^2\over 3 \Delta^2}+{11\over 45}a^4+O(\Delta^2)\right).\label{Dexp}
\end{eqnarray}
Again, one finds that a non-zero ``vacuum energy", $11e^2 a^4/90$,
survives after the $\Delta^{-4}$ and $\Delta^{-2}$ singularities
%(corresponding to static and radiation fields)
are subtracted from Eq.$(\ref{Dexp})$. Note that ``vacuum" does not
mean that EM field vanish in space. As Born's solution appears, the
radiation of the UAC has infinite wave-length, which corresponds to
static fields\cite{schw}.

The singularities in $f(\Delta)=[\sinh (a\Delta/2)]^{-4}$ can also
be removed as follows. First we perform a Fourier transform
\begin{equation}
  \tilde{f}(k) = \int d\Delta e^{ik\Delta}f(\Delta+i\epsilon),
\end{equation}
where $\epsilon$ is a small positive number put by hand to avoid the
singularity. Let the contour to be the one surrounding the upper(lower)
complex $\Delta$-plane for positive(negative) $k$. Since $f(\Delta)$
has periodic singularities at $\Delta=i2n\pi/a$ $(n\in Z)$, above
integration becomes ($\omega\equiv |k|/c$)
\begin{eqnarray}
  \tilde{f}(\omega )&=& \sum_n e^{-2n\pi kc/a} \oint_{C_n}
    d\Delta {e^{ik\Delta}\over \sinh^4(a\Delta/2)}\nonumber\\
  &=& {\pi\over 3}{1\over e^{2\pi\omega c/a}-1}\left({c\omega\over a^2}+
    {(c\omega )^3\over a^4}\right),\label{spec}
\end{eqnarray}
where $\sum_n\equiv\sum_{n=1}^\infty$ for $k>0$, and $\sum_n\equiv
\sum_{n=-\infty}^0$ for $k\le 0$. Then the renormalized $\cal E$
is given by an inverse transform
\begin{eqnarray}
  {\cal E}_{\rm Ren} &=& \lim_{\Delta\to 0} {e^2 a^4\over 2}
    \int_{-\infty}^{\infty}{dk\over 2\pi}e^{-ik\Delta}\tilde{f}(\omega)
  \nonumber\\
    &=&{8 e^2 a^4\over 3}\left[ \Gamma(2)\zeta(2)\left({k_B Tc\over\hbar a}
    \right)^2 + \Gamma(4)\zeta(4)\left({k_B Tc\over\hbar a}\right)^4\right]
    \nonumber\\ &=& {11\over 90}e^2 a^4,
\end{eqnarray}
which is exactly the $O(\Delta^0)$ term in Eq.$(\ref{Dexp})$.

As long as we identify the $\tilde{f}(\omega )$ in Eq.$(\ref{spec})$ to
Planckian-like spectrum, the well-known thermal character with temperature
$T=\hbar a/2\pi c k_B$ arises, though $\tilde{f}(\omega )$ is
different from the ones for isotropic $(3+1)$ dimensional finite
temperature systems by a numerical factor as well as an $\omega^1$ term.
However, above calculation shows that the temperature is simply a dummy
parameter in the mode integration for the renormalized energy. If one
perform a scale transformation $\Delta\to b\Delta$, the temperature in
power spectrum $(\ref{spec})$ becomes $b\hbar a/2\pi$, while the vacuum
energy after integration ($\sim\Delta^0$) is still the same. Both
the temperature and the $\hbar$ in temperature are obtained simply
by extracting Planck factor from Eq.$(\ref{spec})$, which is not
necessary in classical electrodynamics.

Hence we may say that the Unruh effect for a UAC is essentially a part
of the electrodynamics. A non-zero ``vacuum energy" with thermal(Planckian
or non-Planckian) spectrum does not imply that the classical UAC really
experiences a thermal background. No additional Brownian motion
for UAC is needed in classical framework, so we can get rid of the
ill-defined thermal equilibrium for a {\it single particle} or charge. What
a detector in a laboratory far away from the classical charge measured is
exactly those the Born's (or Bondi and Gold's) solution Eq.$(\ref{bornEz})
$-$(\ref{bornBp})$ describes, namely, asymptotic Larmor
radiation\cite{rohr}. This might explain why the predictions of some
proposals given earlier\cite{bell} cannot be distinguished to the results
from quantum electrodynamics.

It should be noticed that the existence of non-zero ``vacuum energy" does
not imply the thermal character. To obtain the Planck factor, one has to
know the global property of the system ($\Delta\in (-\infty,\infty)$)
to make the Fourier transformation, while the ``vacuum energy" is
determined by local field configuration around the charge.
A non-uniformly accelerating charge can also recognize the same ``vacuum
energy" as the ones for UAC at some instants if their accelerations
are the same in that period of time.

Technically, the origin of the non-zero ``vacuum energy" is the non-linear
relations between the expansion variables when the acceleration is not
zero. The retarded(advanced) field strength measured at $x^\mu$ due to the
charge at the point $z^\mu(\tau)$ reads\cite{rohr}
\begin{equation}
  F^{\mu\nu}_{\rm ret(adv)}(x)={e\over r^2c}v^{[\mu}u^{\nu ]}+
    {e\over rc^2}\left\{ c^{-1} a^{[\mu}v^{\nu ]}-u^{[\mu}
    \left( v^{\nu]}c^{-1}a_\alpha u^\alpha \pm a^{\nu ]}\right)\right\},
\end{equation}
where $A^{[\mu}B^{\nu ]}\equiv A^\mu B^\nu -A^\nu B^\mu$, $a^\mu$ is the
four-acceleration of the charge at $z^\mu$, the spacelike vector $u^\mu$
and the scalar propagating length $r$ are defined by
\begin{equation}
  R^\mu_{\rm ret(adv)} \equiv x^\mu -z^\mu(\tau) = r(u^\mu \pm v^\mu/c).
\end{equation}
While the $r$-expansion of $F^{\mu\nu}$ has $r^{-2}$ and $r^{-1}$ terms
only, the $\Delta$-expansion of $F^{\mu\nu}$ for the observer in trajectory
$(\ref{obser})$ has higher order terms because here $r^{-1}=a/\sinh
(a\Delta /2)$ is non-linear. Note that both the static part ($r^{-2}$-term)
and the radiation part ($r^{-1}$-term) contribute to the ``vacuum energy"
in this case.

To conclude, we have another interesting point of view as a remark: the
detector and the field should be considered as a whole, and this problem
is a boundary condition issue\cite{rohr}.  Reversing the direction of
deduction, we may say that the particle recognizes a constant non-zero
``vacuum energy" or field strength at its position because we force the
particle on the track of a hyperbolic motion by choosing $F^{\mu\nu}_{\rm
out}-F^{\mu\nu}_{\rm in}=F^{\mu\nu}_{\rm ret}-F^{\mu\nu}_{\rm adv}=0$
around the charge. This corresponds to some boundary conditions at
infinity. The incoming radiation from past infinity associated with this
particular choice of boundary conditions serves a support to keep the
detector in the hyperbolic trajectory while it's energy dissipates by
radiation. One has, of course, the freedom to choose other boundary
conditions which yield non-uniform accelerations with radiation damping.
Nevertheless, whether there exists such cases depends on the existences of
proper solutions satisfying Maxwell equations as well as Lorentz-Dirac
equations in these particular boundary conditions.

Above viewpoints can be applied to the black hole radiation. One
interprets the point-like detector in Unruh's model sees a ``thermal
energy" simply because the correlation function or the renormalized energy
has a Planckian-like spectrum for some variables when boundary conditions
were chosen. This suggests that, for a black hole in a pure gravity
system, one has to choose some particular boundary conditions for field
equations as well as the equation of motion for the detector to keep the
detector in a rest (hence non-inertial or accelerating) frame relative to
the black hole. Then the information about black hole radiation was encoded
in the gravitational field configuration near the detector, if this static
solution with respect to the clock of the detector exists. However, if
there does not exist proper solution for any boundary condition, then the
thermal radiation is physically meaningless for gravitational detectors
of this type.

\begin{acknowledgments}
I would like to thank J. M. Nester, C. I Kuo, and T. Hirayama for helpful
discussions and comments.
\end{acknowledgments}

\end{document}